# Discovering Latent Concepts and Exploiting Ontological Features for Semantic Text Search

**Vuong M. Ngo** and **Tru H. Cao**

**Abstract**

Named entities and WordNet words are important in defining the content of a text in which they occur. Named entities have ontological features, namely, their aliases, classes, and identifiers. WordNet words also have ontological features, namely, their synonyms, hypernyms, hyponyms, and senses. Those features of concepts may be hidden from their textual appearance. Besides, there are related concepts that do not appear in a query, but can bring out the meaning of the query if they are added. The traditional constrained spreading activation algorithms use all relations of a node in the network that will add unsuitable information into the query. Meanwhile, we only use relations represented in the query. We propose an ontology-based generalized Vector Space Model to semantic text search. It discovers relevant latent concepts in a query by relation constrained spreading activation. Besides, to represent a word having more than one possible direct sense, it combines the most specific common hypernym of the remaining undisambiguated multi-senses with the form of the word. Experiments on a benchmark dataset in terms of the MAP measure for the retrieval performance show that our model is 41.9% and 29.3% better than the purely keyword-based model and the traditional constrained spreading activation model, respectively.

## 1. Introduction

With rapid development of the World Wide Web and e-societies, Information Retrieval (IR) has many challenges in discovering and exploiting those rich and huge information resources. Semantic search improves search precision and recall by understanding user's intent and the contextual meaning of concepts in documents and queries (Huston and Croft, 2010; Losada, et al, 2010; Egozi, et al, 2011).

Concepts are named entities or WordNet words (unnamed entities). Named entities are those that are referred to by names such as people, organizations, and locations (Sekine, 2004) and could be described in ontologies. Each fully recognized named entity (NE) has three features, namely, name, class, and identifier. WordNet words are words in a lexical database (e.g. WordNet database). Each fully recognized WordNet word (WW) has three features, namely, form, direct hypernym, and sense.

Lexical search is not adequate to represent the semantics of queries referring to NEs or WWs. Some examples of NE-based queries are: (1) Search for documents about "*football clubs*"; (2) Search for documents about "*Barcelona*"; (3) Search for documents about "*Paris City*"; (4) Search for documents about "*Paris City, Texas, USA*". In fact, the first query searches for documents containing NEs of the class *Football Club*, e.g. *Chelsea* or *Barcelona*, rather than those containing the keywords "*football club*". For the second query, target documents may mention *Football Club Barcelona* under other names, i.e., the football club's aliases, such as *Football Club Barca*. Besides, documents containing *Barcelona City* or *Barcelona University* are also suitable. In the third query, users do not expect to receive answer documents about entities that are also named "*Paris*", e.g. the actress *Paris Hilton* or *University of Paris* but are not cities. Meanwhile, the fourth query requests documents about a precisely identified named entity, i.e., the *Paris City in Texas, USA*, not the one in France. That are, entity aliases, classes, and identifiers have to be taken into account.

Some examples about WW-based queries are: (1) Search for documents about "*movement*"; (2) Search for documents about "*movement belonging to change*"; and (3) Search for documents about "*movement belonging to the act of changing location from one place to another*". That is because the word *movement* has many different senses. In fact, the first query searches for documents containing not only the word *movement* but also its synonyms, e.g. *motio*n, *front*, *campaign*, and *trend*, or its hypernyms, e.g. *change*, *occurrence*, *social group*, *venture*, and *disposition*. For the second query, users do not expect to receive answer documents about words that are also labelled "*movement*", e.g. *movement belonging to a natural event* and *movement belonging to a venture*, but do not express changes. Meanwhile, the third query requests documents about a precisely identified word sense. The word *movement* means not only the action of changing something but also the act of changing location from one place to another, e.g. *the movement of people from the farms to the cities*.

Moreover, queries may contain both named entities and WordNet words. Some examples of NE-WW based queries are "*temblor in USA*" or *"natural calamity in USA"*, for which documents about "*earthquake in United States of America*" are truly relevant answers.

Besides, there are latent concepts that do not appear in queries but present user's intent. Intuitively, adding correct related concepts to a query will increase the recall and the precision of searching. In contrast, adding incorrect related concepts will decrease performance of IR system. For examples, consider the following queries: (1) Search for documents about "*cities that are tourist destinations of Thailand*"; (2) Search for documents about "*tsunami in Southeast Asia*"; (3) Search for documents about "*settlements are built in west of Jerusalem*"; and (4) Search for documents about "*Barack Obama uses high-tech defences*". For the first query, Chiang Mai and Phuket should be added into the query, because they belong to class *City* and are *tourist destinations of Thailand*. For the second query, countries having relation "*is part of*" with *Southeast Asia* in the exploited ontology should be added into the query, e.g. *Indonesia* or *Philippine*. However, added countries should be those that were actually hit by at least one tsunami, according to the given ontology. So, *Laos* should not be added into the query. For the third query, if there are facts that *settlements are built in* the locations in the *west of Jerusalem*, e.g. *Givat Zeev* and *Pisgat Zeev*, then those locations should add into the query. For the fourth query, bullet-resistant suit should be added into the query; because it *is hyponym of high-tech defences* and the President *Barack Obama have used* a bullet-resistant suit.

In this paper, we propose a new ontology-based text search model with two key ideas as our two contributions. First, it exploits different ontological features of NE and WW existing in documents and queries. Until now, there is no other text search model that formally exploits and presents in documents and queries all above-mentioned NE features or all above-mentioned WW features. Specifically, in a context, after a disambiguation process, if a WordNet word has more than one sense with the equally highest rank, then the most specific common hypernym (msc_hypernym) of those senses will be chosen and the word will be represented by the pair of that hypernym and the form of the word. Meanwhile, other WordNet-based text search models choose one of those senses randomly or all of the senses (Vooheres, 1994; Liu, et al., 2004; Zaihrayeu, et al., 2007; Hsu, et al., 2008; Giunchiglia, et al., 2009). Second, our model expands a query by latent concepts relating to concepts and relations in the original query as asserted in employed ontologies. Our proposal is more general than Fu, et al. (2005), which considered only spatial relations.

In the next section, we discuss related works. Section 3 describes the proposed system architecture and detailed model. Section 4 presents evaluation of the proposed model and discussion on experiment results in comparison to other models. Finally, section 5 gives some concluding remarks.

## 2 Related Works
### 2.1 Exploiting Named Entities

There are works exploiting NEs but not for document search. The Falcons system described in Cheng, et al. (2008) is assisted by users to determine clearly the meaning of queries. In Cheng, et al. (2007), the authors use classes of NEs associated with keywords in a query. However, they are for entity search.

Vallet and Zaragoza (2008), Santos, et al. (2010), Demartini, et al. (2010), and Kaptein, et al. (2010) use only names and classes of NEs, and they are for entity ranking (Balog, et al. 2009).

Gupta and Ratinov (2008), Chang, et al. (2008), Wang, et al. (2009), and Jing, et al. (2010) use only labels of concepts (NE names or WW forms) to represent documents and queries. Moreover, they are for document classification, not document search.

There are some papers using named entity for document search. Bast et al. (2007) considers only entity classes in combination with keywords. In Ahn, et al. (2010), the NameSieve system uses only names and classes of NEs, and limits in four entity class: who, where, when and what. Beside, the system is helped by users to determine clearly the meaning of queries. In Egozi, et al. (2011), the authors use only names of concepts to present documents and queries.

## 2.2 Exploiting WordNet

Voorhees (1994), Liu, et al. (2004) and Hsu, et al. (2008) use all forms of a sense and all forms of every hyponym of a sense in a query. Meanwhile, Zaihrayeu, et al. (2007) uses all forms of a sense to expand a document, and Wang, et. al. (2004) and Giunchiglia, et al. (2009) additionally use all forms of every hypernym of a sense in a document. Mihalcea and Moldovan (2000) use senses in both queries and documents, and all forms of every hypernym of a sense in a document.

Moreover, since the above-surveyed papers, except for Mihalcea and Moldovan (2000), use word forms to represent word senses, it may reduce the precision of system. Indeed, a query containing a word having form $f$ and sense $x$ could also match to documents containing a word having the same form $f$ but different sense $y$. The drawback is similar with using only word forms of hypernyms and hyponyms of senses.

Especially, in case a word has more than one sense determined by a Word Sense Diambiguation (WSD) algorithm, the above works choose randomly one sense from those senses, which may decrease the retrieval performance if that is a wrong choice. In contrast, in our system, such a word is represented by the combination of its form and the most specific common hypernym of the senses.

## 2.3 Exploiting Latent Concepts

Some systems improve document retrieval performance by expanding queries with user's participation, such as Sanderson (2004), Balog, et al. (2008), Castellani, et al. (2009), Meij, et al. (2009) and Ahn, et al. (2010). Whereas, Bendersky and Croft (2008), and Huston and Croft (2010) identify key concepts in queries to remove unimportant words.

In Wang and Zhai (2008), the authors exploit synonyms or co-occurring relations in search engine logs for repairing or expanding queries. In Losada, et al. (2010), the system uses pseudo- relevance feedback to expand queries. However, the two systems do not take account relations in a query.

In Tran, et al. (2007), the authors map concepts of a query to an ontology to find suitable related concepts. In Cheng, et al. (2007), the target problem is to search for named entities of specified classes associated with keywords in a query. Different from our model, the two systems do not take account relations in queries and they are for question-and-answering but not document search.

In Castells, et al. (2007), the system finds identified named entities belonging to a class of NE in a query, after the query's vector is constructed by the NEs. This step is unnecessarily time consuming. In our proposed models, the query and document vectors having the entity class can be constructed and matched right away. Beside, its queries must be specified by RDQL. Similarity, in Kasneci, et al. (2008), queries must be written by SPARQL. Concepts and relations must be clearly specified by users. Whereas, this need not in our system. Moreover, the work is for question-and-answering, not document retrieval.

Spreading Activation (SA) is a popular algorithm for query expansion. But pure-SA would return most results irrelevant to queries (Berthold, et al., 2009). So, SA algorithms have been constrained by some methods to improve retrieval performance.

In Rocha, et al. (2004), the authors propose a hybrid spread activation algorithm that combined SA algorithm together with ontology based information retrieval. In Aswath, et al. (2005), the system uses a two-level SA network to activate strongly positive and strongly negative matches based on keyword search results.

In Schumacher, et al. (2008), the system finds answers of given query and added into the query before using an SA algorithm. Besides, Hsu, et al (2008) expands query by using SA on all relations in WordNet and only selecting kernel words that are activated and represent the content of a query by some rules.

In Jiang and Tan (2009), the authors map the original query to a keyword set and searches for documents relating to the keyword set. After that, the documents are pre-annotated with information of an ontology and the initial concepts are extracted from the retrieved documents. An SA algorithm is used to find concepts semantically relating to the concepts in the ontology. Finally, the activated concepts are used to re-rank the documents to present for user. In Lee, et al. (2010), the system sets up an associative network with nodes being web pages and links between the nodes being relations between the web pages. Initial nodes of SA algorithm are web pages that are strongly associated to given query. Next, other nodes (web pages) of their network are activated.

However, the above Constrained-SA (CSA) models do not use relations in a given query to constrain spreading. Meanwhile, our relation-

CSA method activates concepts relating to concepts and relations in queries. In Fu, et al. (2005), the authors use the relations in a query to expand the query. However, the work only exploits spatial relations (e.g. near, inside, north of). In contrast, in this paper, we propose more general rules for query expansion.

## 3 Ontology-based Text Search
### 3.1 System Architecture

Our proposed system architecture of semantic text search is shown in Figure 1. It has two main parts. Part 1 presents document and query annotation and expansion. Part 2 presents the query expansion module using a relation-CSA (RCSA) method.

Our proposed model needs an ontology having: (1) a comprehensive class catalog with a large concept population for expressing clearly information of documents and queries; and (2) a comprehensive set of relations between concepts and facts for expanding queries with latently related concepts. Since no single ontology is rich enough for every domain and application, merging or combining multiple ontologies are reasonable solutions (Choi, et al. 2006). So we have combined 3 ontologies, namely, KIM, WordNet, and YAGO to have a rich ontology for our model.

In this work we employ KIM (Kiryakov, et al. 2005) for automatic NE recognition and semantic annotation of documents and queries. The KIM PROTON ontology contains about 300 classes and 100 attributes and relations. KIM World Knowledge Base (KB) contains about 77,500 entities with more than 110,000 aliases. NE descriptions are stored in an RDF(S) repository. Each NE has information about its specific class, aliases, and attributes (i.e., its own properties or relations with other NEs). The average precision and recall of the NE recognition engine are about 90% and 86%, respectively[1].

WordNet (Fellbaum, 1998) is a lexical database for English organized in synonym sets (synsets). There are various semantic relations between these synonym sets, such as hypernym, hyponym, holonym, meronym, and similarity. WordNet version 3.0 contains about 155,000 words organized in over 117,000 synsets.

Since KIM ontology and WordNet define only a small number of relations, and KIM KB contains a limited number of facts, we employ YAGO (Yet Another Great Ontology) (Suchanek, et al. 2007; Suchanek, et al. 2008) for an ontology of relations in the system. It contains about 1.95 millions entities, 93 different relation types, and 19 millions facts about specific relations between entities. The correctness of the facts is about 95%. In addition, with logical extraction techniques and a flexible architecture, YAGO can be further extended in future. Note that, to have more relation types and facts for experiments, we can manually combine it with Wikipedia. We use KIM, WordNet, YAGO as the NE, WW, and Fact ontologies in our system, respectively.

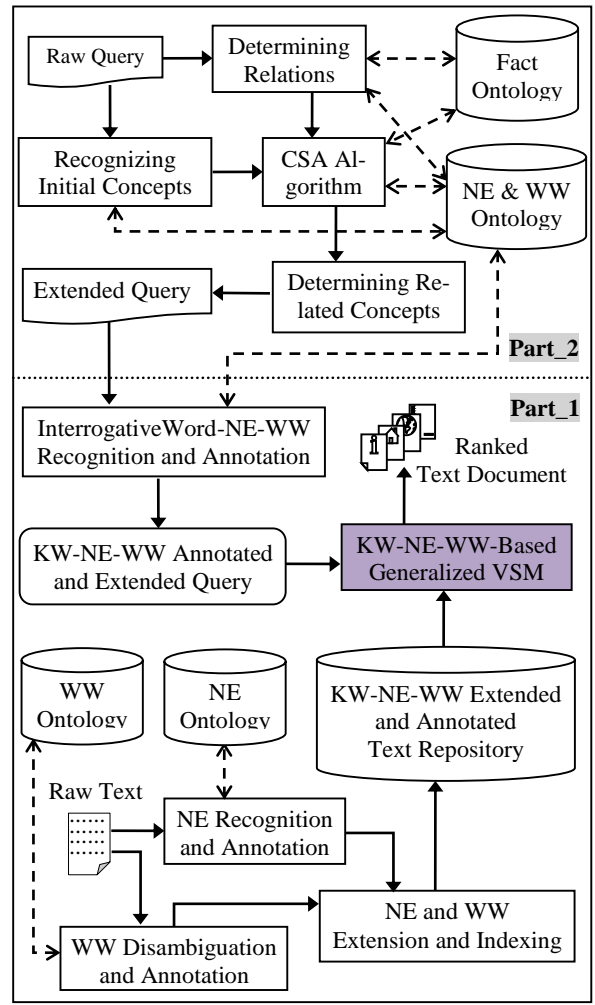

Figure 1. System architecture for semantic search

The NE Recognition-and-Annotation module and WW Disambiguation-and-Annotation module extract and embed NE features and WW features in a raw text, respectively. The text is then indexed by contained NE features, WW features, and keywords, and stored in the Extended KW-NE-WW Annotated Text Repository. Meanwhile, the InterrogativeWord-NE-WW Recognition-and-Annotation module extracts and embeds the most specific NE features and WW features in the extended query, and replaces the interrogative word if existing by a suitable class. Semantic document search is performed via the KW-NE-WW-Based Generalized Vector Space Model (VSM) module.

---
[1] It is reported at http://www.ontotext.com/kim/performance.html.

## 3.2 Word Sense Disambiguation

To choose the intended sense of a word in a context, a WSD algorithm is employed. Supervised WSD systems have high accuracy (Pradhan, et al. 2007) but need manually sense-tagged corpora for training. In IR, training corpora of a supervised WSD algorithm need to be large which are usually laborious and expensive to construct. Knowledge-based WSD systems (Liu, et al. 2005; Sinha and Mihalcea, 2007; Navigli and Lapata, 2007; Agirre and Soroa, 2009a) are developed to overcome the knowledge acquisition bottleneck and avoid manual effort. Besides, for specific domains, knowledge-based WSD systems have better performance than generic supervised WSD systems trained on balanced corpora (Agirre, et al. 2009b). We use Personalizing PageRank algorithm of Agirre and Soroa (2009a) having 56.8% accuracy for our WordNet based WSD. Moreover, we enhance it by using Pos-Tagger and Lemmatization in Toutanova, et al. (2003) having 97.24% accuracy. However, if a word has two or more probable senses, then our WSD algorithm will choose the most specific common hypernym of the senses in hypernym hierarchy of WordNet. We use WordNet version 3.0 for the WSD algorithm. Figure 2 describes the difference between the traditional KB-based WSDs and our KB-based WSD.

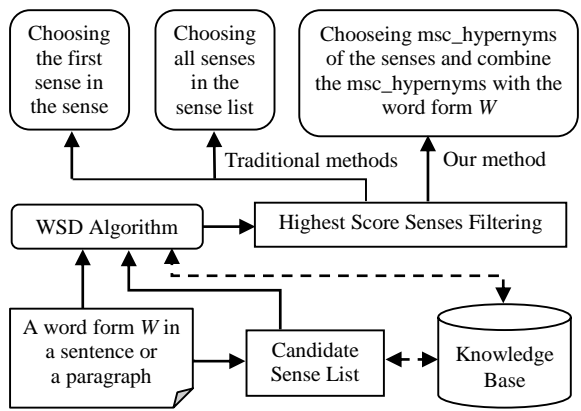

Figure 2. Difference between the traditional KB-based WSDs and our KB-based WSD

## 3.3 Annotating and Expanding in Queries and Documents

We propose a generalized VSM in which a document or a query is represented by a vector over a space of generalized terms. Each term is a NE feature, a WW feature, or a keyword. As usual, similarity of a document and a query is defined by the cosine of the angle between their representing vectors. Our work has implemented the model by developing a platform modified from Lucene[2]. The system automatically processes documents for KW-NE-WW-based searching in the following steps:

1. Removing stop-words in the documents.
2. Recognizing and annotating NEs in the documents using KIM[3].
3. Disambiguating and annotating WWs that are not NEs in the document using the WSD algorithm mentioned in section 3.2.
4. Words not defined in KIM and WordNet are treated as plain keywords.
5. Extending the documents with implied NE features. That is, for each entity named $n$ possibly with class $c$ and identifier $id$ in a document, the triples $(n/*/*)$, $(*/c/*)$, $(n/c/*)$, $(alias(n)/*/*)$, $(*/super(c)/*)$, $(n/super(c)/*)$, $(alias(n)/c/*)$, $(alias(n)/super(c)/*)$, and $(*/*/id)$ are virtually added to the document. Here $alias(n)$, $super(c)$, $syn(w)$ and $super(h)$ respectively denote any alias of $n$, any super class of $c$, any synonym of $w$, and any super hypernym of $h$ in the ontology and knowledge base of discourse.
6. Extending the document with implied WW features:
   - If the sense $s$ of the word is determined, then $s$ and its expanded features $form(s)$, $hypernym(s)$, $form(hypernym(s))$, $form(s)/hypernym(s)$ are added into the document.
   - If the word has more than one sense with $f$ and $msc\_hypernym(possible\_senses(f))$ as its apparent form and the most specific common hypernym, respectively, then $f$ and $f/msc\_hypernym(possible\_senses(f))$ and their expanded features:
   
   $form(msc\_hypernym(possible\_senses(f)))$,
   
   $msc\_hypernym(possible\_senses(f))$,
   
   $form(hypernym(msc\_hypernym(possible\_senses(f))))$,
   
   $hypernym(msc\_hypernym(possible\_senses(f)))$,

---

*f/hypernym*(*msc_hypernym*(*possible_senses*(*f*))) are virtually added to the document.

7. Original and implied features of NE and WW, and plain keywords are indexed by S-Lucene.

A query is also automatically processed in the following steps:

1. Removing stop-words in the query.
2. Recognizing and annotating NEs in the query.
3. Disambiguating and annotating WWs that are not NEs in the query.
4. Words not defined in KIM and WordNet are treated as plain keywords.
5. Representing each recognized entity named *n* possibly with class *c* and identifier *id* by the most specific and available triple among (*n*/*/*), (*/c*/*), (*n*/*c*/*), and (*/*/*id*).
6. Representing each recognized WordNet word:
   - If the sense *s* of the word is determined, then the word is represented by *s*.
   - If the word has more than one sense with *f* and *msc_hypernym*(*possible_senses*(*f*)) as its apparent form and the most specific common hypernym, respectively, then the word is represented by *f/msc_hypernym* (*possible_senses*(*f*)).

Besides, there is latent information of the interrogative words *Who*, *What*, *Which*, *When*, *Where*, or *How* in a query. For example, given the query "*Where was George Washington born?*", the important terms are not only the NE *George Washington* and the WW "*born*", but also the interrogative word *Where*, which is to search for locations or documents mentioning them. For instance, *Where* in this example should be mapped to the class *Location* of NE. The mapping could be automatically done with high accuracy using the method proposed in (Cao, et al. 2008).

### 3.4 Discovering Latent Concepts in Queries

The followings are the six main steps of our RCSA method to determine relevant latent related concepts for a query:

1. Recognizing relation phrases: Relation phrases are prepositions, verbs, and other phrases representing relations, such as *in*, *on*, *is*, *near*, *north of*, *live in*, *located near*, *was actress in*, *is author of*, and *was born*. We have implemented a relation phrase recognition using the ANNIE tool of GATE (Cunningham, et al. 2006).
2. Determining relations: Each relation phrase recognized in step 1 is mapped to the corresponding relation in fact ontology or NE ontology by a manually built dictionary. For example, "*was actress in*" is mapped to *actedIn*, "*is author of*" is mapped to *wrote*, and "*nationality is*" is mapped to *isCitizenOf*.
3. Recognizing initial concepts: we find concepts in the query by mapping the words expressed in the query to entity names or word forms in the exploited ontologies. These are original concepts in the query and initial concepts of the method.
4. Presenting each relation in the query in the form $C_1RC_2$, where R is a relation found in step 2, and $C_1$ and $C_2$ are initial concepts found in step 3.
5. Determining related concepts. Let $C_4$ be a latent concept derived from a relation $C_1RC_2$.
   - If $C_2$ is a NE having identifier and belonging to class *Location*:
     o If R is described by a verb and a spatial relation phrase, e.g. "born in the north of", find $C_4$ that satisfies $C_4R_SC_2$ in the employed NE ontology and $C_1R_FC_4$ in the Fact ontology, where $R_S$ is the relation expressed by the spatial relation phrase and $R_F$ is the relation expressed by the verb.
     o Otherwise, find $C_4$ that satisfies $C_4$ *is_part_of* $C_2$ in the NE ontology and $C_1RC_4$ in the Fact ontology.
   - If $C_2$ is a NE class only, find $C_4$ that satisfies $C_4$ *is_subClass_of* $C_2$ in the NE ontology and $C_1RC_4$ in the Fact ontology.
   - If $C_2$ is a WW, find $C_4$ that satisfies $C_4$ *is_hyponym_of* $C_2$ in the WW ontology and $C_1RC_4$ in the Fact ontology.
6. Before being added into the query, the latent concepts are represented by their main entity aliases or word forms.

Comparing with pure-SA algorithm, the RCSA algorithm has two constraints as follows: (1) distance constraint: only concepts having direct relations, in accordance to the exploited ontology, with original nodes in queries are activated; and (2) relation constraint: relations used for spreading in the Fact ontology must appear in the query.

For the computational cost, we note that document annotation is performed offline, while queries are typically short and thus query annotation and expansion could be done quickly. Therefore, the query answering time is not a problem.

## 4 Experiments

Evaluation of a retrieval method requires two components being a test dataset and quality measures (Baeza-Yates and Ribeiro-Neto, 1999; Manning, et al. 2008). The L.A. Times document collection is employed, which was used by 15 papers among the 33 full-papers of SIGIR-2007 and SIGIR-2008 about text IR using TREC dataset. The L.A. Times consists of more than 130,000 documents in nearly 500MB. Next, queries in the QA Track-1999, which have answer documents in this document collection, are used. So, there are 124 queries of 200 queries in this Track chosen.

Table 1. MAPs and two-sided p-values of the Lexical, NE+KW, WW+KW and NE+WW+KW models.

| Model A and MAP | Model B and MAP | Improvement | Two-Sided P-Value |
|---|---|---|---|
| NE+WW +KW **0.6024** | Lexical **0.5099** | 18.1% | 0.02004 |
| | NE+KW **0.5652** | 6.6% | 0.03359 |
| | WW+KW **0.5391** | 11.7% | 0.04118 |

Table 2. MAPs and two-sided p-values of the Lexical, CSA and RCSA models.

| Model A and MAP | Model B and MAP | Improvement | Two-Sided P-Value |
|---|---|---|---|
| RCSA **0.6594** | Lexical **0.5099** | 29.3% | 0.02952 |
| | CSA **0.5592** | 17.9% | 0.04987 |

We have evaluated and compared the IR models in average Precision-Recall (P-R) curves, average F-measure-Recall (F-R) curves, and mean average precision (MAP) values (Baeza-Yates and Ribeiro-Neto, 1999; Manning, et al. 2008). Because, average P-R curves and average F-R curves represent commonly the retrieval performance and allow comparison of those of different systems. The closer the curve is to the right top corner, the better performance it represents (Manning, et al. 2008). Whereas, MAP is a single measure of retrieval quality across recall levels and considered as a standard measure in the TREC community (Voorhees and Harman, 2005). Obtained values of the measures presented above might occur by chance. Therefore, a statistical significance test is required (Hull, 1993). We use Fisher's randomization (permutation) test for evaluating the significance of the observed difference between two systems, as recommendation of Smucker, et al. (2007). As shown Smucker, et al. (2007), 100,000 permutations were acceptable for a randomization test and the threshold 0.05 of the two-sided significance level, or two-sided p-value, could detect significance.

We conduct experiments to compare the results obtained by the following seven different search models:
1. Lexical: This is the Lucene text search engine as a tweak of the traditional keyword-based VSM.
2. NE+KW: This is the model only exploiting features of NEs to annotate and expand documents and queries.
3. WW+KW: This is the model only exploiting features of WW to annotate and expand documents and queries.
4. NE+WW+KW: This is the model combining NE+KW and WW+KW, as presented in section 3.3.
5. CSA: This is the model using the traditional constrained SA algorithm. It expands queries by broadcasting all direct-links to original concepts in the Fact ontology to find related concepts. The expanded queries and documents of the CSA model are represented by keywords.
6. RCSA (6): This is the model improving the above CSA model. The RCSA model only uses links presented in a query to find related concepts, as presented in section 3.4.
7. Semantic Search: This is the model combining RCSA and NE+WW+KW, as presented in section 3.

The MAP values of the models and two-sided p-values of randomization tests between them in Table 1 show that taking into account ontological features in queries and documents does enhance text retrieval performance; NE+WW+KW performs about 18.1%, 6.6%, and 11.7% better than the Lexical, NE+KW and WW+KW models in terms of the MAP measure, respectively.

In Table 2, we see that RCSA model really performs about 29.3% and 17.9% better than the Lexical and CSA models in terms of the MAP measure, respectively. So, discovering latent concepts in a query does enhance text retrieval performance.

Finally, Table 3 and Figure 3 show that text retrieval performance is improved by the combination of discovering latent concepts and exploiting logical feature in documents and queries. In terms of the MAP measure, Semantic Search performs about 41.9% and 29.3% better than the Lexical and CSA models, respectively. Beside, Semantic Search also performs about 28%, 34.2%, 20.1%, and 9.7% better than the NE+KW, WW+KW, NE+WW+KW and RCSA models, respectively.

Table 3. MAPs and two-sided p-values of the Semantic Search model and the other six models.

| Model A and **MAP** | Model B and **MAP** | Improvement | Two-Sided P-Value |
|---|---|---|---|
| **Semantic Search 0.7233** | Lexical **0.5099** | 41.9% | 0.01071 |
| | NE+KW **0.5652** | 28.0% | 0.00313 |
| | WW+KW **0.5391** | 34.2% | 0.00845 |
| | NE+WW+KW **0.6024** | 20.1% | 0.01791 |
| | CSA **0.5592** | 29.3% | 0.01255 |
| | RCSA **0.6594** | 9.7% | 0.04516 |

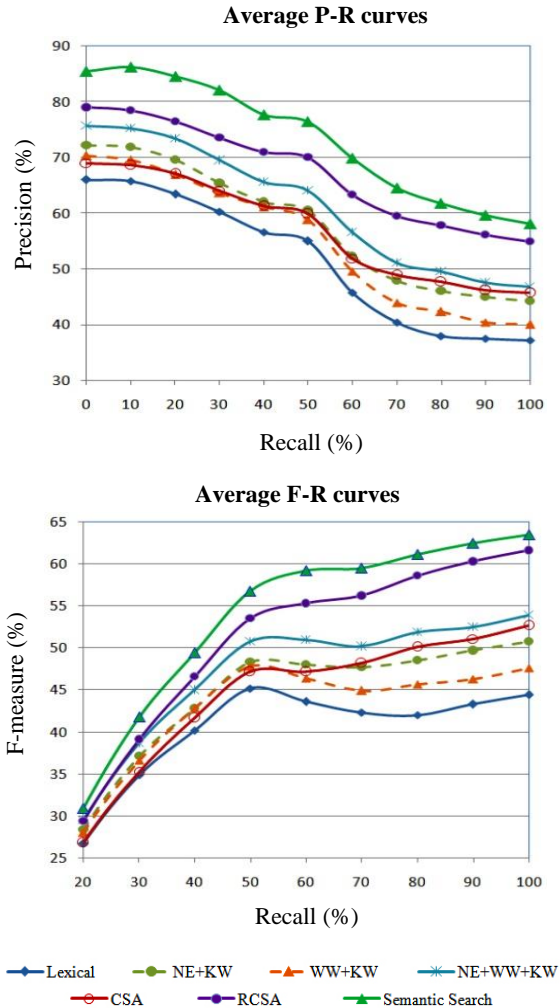

Figure 3. Average P-R and F-R curves of the seven search models on 124 queries of TREC

## 5  Conclusion

We have presented the generalized VSM that exploits and annotates ontological features of named entities and WordNet words in documents and queries for semantic text search. In case a word has more than one sense determined by a WSD algorithm, the word is represented by the combination of its form and the most specific common hypernym of those senses. Besides, our model expands a query by discovering relevant latent concepts in the query by constrained spreading activation using relations in the query.

The conducted experiments on a TREC dataset have showed that our semantic search improves the search quality in terms of the precision, recall, F, and MAP measures. Although this work uses VSM for proving the advantage of exploiting the proposed ontological features and discovering latent concepts in text search, it could be adapted for other information retrieval models as well.